%% file: main.tex
\documentclass[copyright,creativecommons]{eptcs}
\providecommand{\event}{AMMSE 2011} 

\usepackage{breakurl}             

\usepackage{amsmath}
\usepackage{amsfonts}
\usepackage{amssymb}
\usepackage{subfig}

\usepackage{url}
\usepackage{times}
\usepackage{graphicx}

\usepackage[latin1]{inputenc}
\usepackage{latexsym}
\usepackage{alltt}
\usepackage{shortvrb}  
\usepackage{dsfont}    

\title{Formal Model Engineering for Embedded Systems Using Real-Time Maude}

\author{Peter Csaba \"Olveczky
\institute{Department  of Informatics, 
           University of Oslo}
}

\def\titlerunning{Formal Model Engineering for Embedded Systems Using Real-Time Maude}
\def\authorrunning{P. C. \"Olveczky}

\begin{document}
\maketitle
 
\input{abstract}
\MakeShortVerb{\@} 
\input{intro}
\input{aadl}

\input{sync-aadl}
\input{ptolemy}

\input{moment2}

\input{emotions}
\input{musab}
\input{concl}

\bibliographystyle{eptcs} 
\bibliography{bibl}

\end{document}

%% file: abstract.tex
\begin{abstract}
This paper motivates why Real-Time Maude should be well suited to provide 
a formal semantics and formal analysis capabilities to modeling languages
for embedded systems. One can then use the code generation facilities
of the tools for the modeling languages to automatically synthesize  Real-Time Maude 
verification models from   design models, enabling  a formal model engineering
process that combines the convenience of modeling
using an 
informal but intuitive modeling language with formal
verification. We give a brief overview six fairly  different modeling 
formalisms for which Real-Time Maude has provided the formal semantics and (possibly) formal analysis.
These models include behavioral subsets of the avionics modeling standard AADL, 
Ptolemy~II discrete-event models, two EMF-based timed model transformation systems, 
and a modeling language  for handset software. 
\end{abstract}

%% file: intro.tex
\section{Introduction}

It is  now acknowledged  that one of the most promising approaches to
inject  formal methods into industrial model-driven engineering is to
endow the domain-specific modeling languages used in
industry with formal verification capabilities. 
This can be achieved by  defining the formal semantics of the 
informal modeling language, and then  automatically synthesize a
\emph{formal verification model} from the informal \emph{design
  model}.  This enables a \emph{formal
  model engineering} process that combines the convenience of modeling
using an 
informal but intuitive modeling language with formal
verification. Furthermore, the synthesis  of the verification model
can 
take advantage of the code generation infrastructure provided by  a number of advanced
modeling tools to support the
generation of deployment code from design models; this infrastructure
can also  be used to generate a formal model.

For the formal model engineering process to be useful in practice, the
target formal language and tool must not only be expressive enough
to define the formal semantics of  the informal language
and provide automated simulation and model checking capabilities, but
should also
\begin{enumerate}
\item allow the users to define formal analysis commands without
 understanding  the formal language or the formal
  representation 
of their models; and
\item provide formal analysis results, such as
  counterexamples
in temporal logic model checking,  that the user can easily
understand.
\end{enumerate} 

Formal model engineering is crucial for real-time embedded
systems (RTESs)---such as automotive, avionics, and 
medical systems---that are hard to design correctly, since subtle
timing aspects impact system functionality, yet are safety-critical
systems whose failures may cause great damage to persons and/or
valuable assets.
There exist a  number of formal analysis tools for real-time systems
(see~\cite{farn-wang-survey} for a survey). However,  there is a
significant gap between the formalisms of these tools, such as timed
automata~\cite{timedautomata} or timed Petri nets~\cite{timedPetri,srba},  
that  sacrifice expressiveness for decidability, and the
expressiveness of modeling languages for industrial RTESs, that were,
after all, designed for modeling convenience. 

In contrast to such formal tools, Real-Time
Maude~\cite{tacas08}---which extends the 
rewriting-logic-based Maude system~\cite{maude-book} to support the
formal specification, simulation, and model checking of real-time
systems---emphasizes expressiveness and ease of specification over
algorithmic decidability of key properties.  In
Real-Time Maude, the state space and the functional properties of the
system are
defined as an   equational specification, the instantaneous
local transitions are modeled as rewrite rules, and time advance is
modeled  by tick rewrite rules. Real-Time Maude is
particularly suitable for modeling real-time systems in an
object-oriented style.
Because of its expressiveness, Real-Time Maude has been successfully
applied to a wide range of advanced state-of-the-art systems that
are beyond the pale 
of timed automata, including 
wireless sensor networks algorithms~\cite{mike-wsn,ogdc-tcs},    
 scheduling algorithms that require unbounded queues~\cite{fase06}, and
large multicast protocols~\cite{aer-journ,norm-paper}.

Real-Time Maude's natural model of time, together with its
expressiveness, should make  it a suitable semantic framework for 
 modeling languages for RTESs; such languages
then also get the following 
formal analysis capabilities  for free: 
\begin{itemize}
\item simulation;
\item reachability analysis;
\item linear temporal logic (LTL) model checking for  for untimed LTL
  properties, as well as time-bounded LTL model checking for analyzing
  systems with an infinite reachable state space; and
\item TCTL${}_{\leq, \geq}$ model checking of
  \emph{timed} CTL formulas~\cite{lepri-tctl}.
\end{itemize}
It is also worth mentioning  that the use of both equations 
(to define the functional aspects of the system) and rewrite rules
(to model its concurrent transitions)  makes 
model checking significantly more efficient than if 
deterministic behaviors were also encoded as rules,
 since only rewrite rules contribute
to the reachable state space. 

The possibility to equationally define \emph{parametric} atomic
state propositions (and ``state patterns'' for reachability
analysis) in Real-Time Maude allows us to predefine  a
useful set of parametric state propositions in the Real-Time Maude interpreter
of a language. These propositions make  it easy for the user to
define his/her search and LTL model checking commands  without having to understand Real-Time
Maude or the formal representation of his/her model, as illustrated in
this paper. 

A key requirement to (i) understand the results of  Real-Time
  Maude analyses, and (ii)   being able to map them back into
  the original modeling formalism is to have, respectively, a small
  \emph{representational distance} between the original models and
  their formal counterparts, and   a one-to-one correspondence
  between these models. Since  hierarchical
  composition  and encapsulation play key roles in industrial modeling
  languages, the possibility of  defining
  \emph{hierarchical} objects enable us to achieve both  goals.

In this extended abstract, I give an overview of six fairly
different modeling languages for RTESs that have a Real-Time
Maude semantics,  and briefly explain how they have been 
endowed  with  Real-Time Maude, or in some cases Maude,
simulation and model checking capabilities. Papers on the semantics of each
modeling have been published
elsewhere~\cite{musab-fase09,icfem11,ptolemy-journal,fase10,fmoods10,rivera-thesis}. 

%% file: aadl.tex
\section{Behavioral AADL}

The  \emph{Architecture Analysis \&{} Design Language}
(AADL)~\cite{aadl} is an industrial standard used in
avionics, aerospace, automotive, medical devices, and robotics
communities to describe a performance-critical embedded real-time
system as an assembly of software components mapped onto an execution
platform.

 In joint work with Jos\'e Meseguer, I
have defined the Real-Time Maude semantics of a subset of the
\emph{software components} of AADL. This subset defines the
architectural and behavioral specification of a system as a set of
hierarchical components with ports and  connections, and with the
behaviors defined by Turing-complete transitions systems. 
The references~\cite{fmoods10,aadl-techrep} 
explain the Real-Time Maude
semantics for the targeted fragment of AADL in detail.
In particular, the semantics of a component-based language can naturally
be defined in an object-oriented style, where each
 component instance is modeled as  an object, and where  the hierarchical structure
of  components is reflected in the nested
structure of objects.

Together with Artur Boronat, we have developed the 
\emph{AADL2Maude}  plug-in  for the 
Open Source AADL Tool Environment (OSATE),
which is a set of Eclipse plug-ins. 
\emph{AADL2Maude}
uses OSATE's code generation
facility to automatically generate Real-Time Maude specifications from
AADL models. However, the Real-Time Maude analysis itself is not yet integrated
into OSATE. 

To allow the user to conveniently define
his/her search and model checking commands without knowing Real-Time
Maude or the 
Real-Time Maude representation of AADL models, we have defined some
useful functions and parametrized state propositions. 
For example, 
the term 

\small
\begin{alltt}
  value of \(v\) in component \(\mathit{fullComponentName}\) in \(\mathit{globalComponent}\)
\end{alltt} 
\normalsize

\noindent gives the value of the state variable 
\(v\) in the thread identified by the full name
\(\mathit{fullComponentName}\) 
 in the system \(\mathit{globalComponent}\).
Likewise, 

\small
\begin{alltt}
  location of component \(\mathit{fullComponentName}\) in \(\mathit{globalComponent}\)
\end{alltt}
\normalsize

\noindent  gives the current location/state in the
transition system  in the given thread.

We have applied the \emph{AADL2Maude} code generator and Real-Time
Maude to formally analyze an AADL model, developed by Min Young Nam
and Lui Sha at UIUC, of a simple  network of medical devices, 
 consisting of a
\emph{controller}, a \emph{ventilator machine} that assists a
patient's breathing during surgery, and an
\emph{X-ray} device~\cite{phuket08}. Whenever a button is pushed to
take an X-ray, the ventilator 
machine should pause for two seconds, starting one second after the
button is pushed, and the X-ray should be taken after two seconds.
The  initial state is 
\texttt{\char123MAIN system Wholesys\! .\! impl\char125}, and 
\texttt{MAIN\! ->\! Xray\! ->\! xmPr\! ->\! xmTh} denotes the
full name of the X-ray machine thread. 
The   
ventilator machine must be  pausing when an X-ray is being taken, so that
the X-ray is not blurred. The following 
 search command 
checks whether  an \emph{undesired} state, where the X-ray thread
@xmTh@ is in local state @xray@ 
 while the ventilator
thread @vmTh@ is \emph{not} in state @paused@, can be reached from the
initial state (the unexpected result shows a concrete  unsafe state
that can be reached from the initial state):

\small
\begin{alltt}
Maude> \emph{(utsearch\! [1]\!} 
         \emph{\texttt{\char123}MAIN system Wholesys\!\! .\!\! impl\texttt{\char125} =>* \texttt{\char123}C:Configuration\texttt{\char125}}
          \emph{such that} 
           \emph{((location of component (MAIN -> Xray -> xmPr -> xmTh)}
               \emph{in C:Configuration) == xray}
            \emph{and} \emph{(location of\! component}\! \emph{(MAIN\! ->\! Ventilator\! ->\! vmPr\! ->\! vmTh)}
                   \emph{in C:Configuration) =/= paused) .)} \vspace{2.6mm}
Solution 1    C:Configuration --> ...
\end{alltt}
\normalsize

For LTL  and timed CTL model checking purposes, our tool has 
pre-defined useful parametric atomic propositions, such as 
\emph{full thread name}\,\,\verb+@+\,\,\emph{location}, which  holds when
the thread is in state \emph{location}, and 
@value of@ \emph{variable} @in component@ \emph{full thread name}
@is@ \emph{value}, which holds if the current value of the local
transition system variable \emph{variable} in the given thread equals
\emph{value}.

We can  use time-bounded LTL  model checking  to verify that 
 an X-ray must be taken
within three seconds of the start of the system (this command returned a counter-example revealing 
a subtle and previously unknown design error):

\small
\begin{alltt}
Maude> \emph{(mc \texttt{\char123}MAIN system Wholesys\! .\! impl\texttt{\char125} |=t}
           \emph{<> ((MAIN -> Xray -> xmPr -> xmTh) @ xray) in time <= 3 .)}\vspace{2.5mm}
Result ModelCheckResult :  counterexample( ... )
\end{alltt}
\normalsize

%% file: sync-aadl.tex
\section{Synchronous AADL}

In a number of systems targeted by AADL, such as integrated
modular avionics systems and distributed control systems in motor
vehicles, the system design is 
essentially a \emph{synchronous design} that must realized in an
asynchronous distributed setting. The key idea of the \emph{PALS architectural
  pattern}~\cite{icfem10,dasc09} is to reduce the design,
verification, and implementation of a distributed real-time system to
that of its much simpler synchronous version, provided
 that the network infrastructure guarantees bounds on the
 messaging delays and the skews of the local clocks. For
 a synchronous design $\mathit{SD}$ and network bounds $\Gamma$,
 we then have a semantically equivalent asynchronous distributed design
 $\mathit{PALS}(\mathit{SD}, \Gamma)$. Not only is this
 important from a 
 system design perspective, but  it also makes model checking verification
 feasible. For example, in~\cite{icfem10}  we show that for the 
 avionics system mentioned below, the synchronous
design model has 185 reachable states and can be model checked in less
than a second in Maude, whereas---even restricted to perfect local clocks and message
delays 0 and 1---the corresponding asynchronous model has more than 3
million reachable states and cannot be model checked in reasonable
time. 

To allow modelers to take advantage of PALS while using AADL,
I have, in joint work with Kyungmin Bae, Jos\'e Meseguer, and Abdullah Al-Nayeem,
 identified a ``synchronous subset'' of behavioral AADL, called
\emph{Synchronous AADL},  that is
suitable to define the synchronous PALS designs in AADL~\cite{icfem11}.

To support formal model engineering, we have integrated the Real-Time Maude analysis of 
 Synchronous AADL models  into OSATE. 
The \emph{SynchAADL2Maude} tool is an OSATE plug-in that uses
OSATE's model traversal facility
to both support the generation of a Real-Time Maude model from an
Synchronous AADL instance model, and for analyzing  the   
synthesized  Real-Time Maude model from \emph{within OSATE}.
\emph{SynchAADL2Maude} inherits the convenient predefined atomic
propositions from our AADL analysis library, and also  allows the
modeler to define  new formulas and LTL model checking commands in
 XML  within OSATE. 

\begin{figure}[htb] \center
\includegraphics[width=130mm]{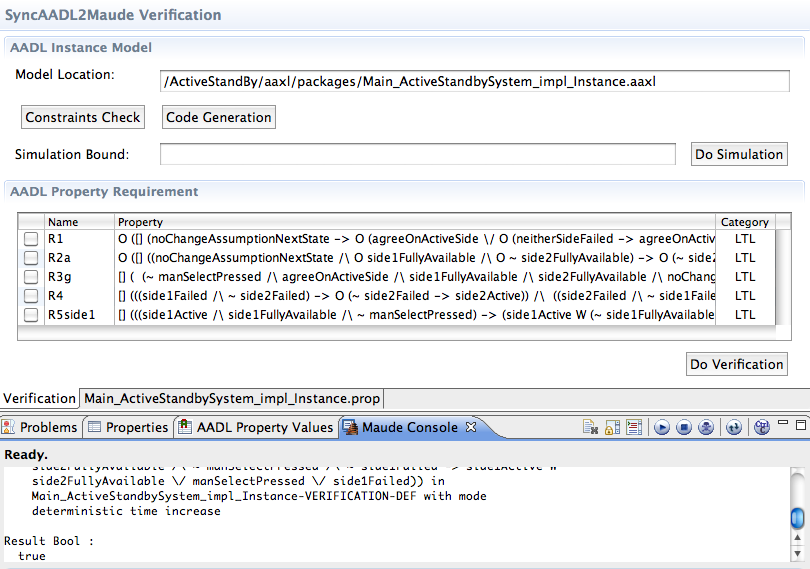}
\caption{\label{fig:syncaadl2maude} SynchAADL2Maude window in OSATE.}
\end{figure}

We have used \emph{SynchAADL2Maude}  to verify a Synchronous AADL model 
of an avionics system based on a specification by Steve Miller
and Darren Cofer at  Rockwell-Collins~\cite{dasc09}. 
Figure~\ref{fig:syncaadl2maude} shows the \emph{SynchAADL2Maude}
window for this  example. Real-Time Maude
code generation  and  model checking are performed by clicking on the 
\texttt{Code Generation} button  and  the \texttt{Do Verification}
button, respectively.
The LTL properties that the avionics system should satisfy have 
been entered into the tool, and are shown in the ``AADL Property
Requirement'' table. 
The \texttt{Do Verification} button has 
been clicked and the results of the model checking are shown in
the ``Maude Console.''

The properties to be verified are managed by the associated XML property
file.
For example, to add an LTL model checking command
to verify a property \texttt{R1}  defined as the LTL formula below, we
just add 
the following \texttt{command} tag to the property file:

\small
\begin{alltt}
<command>   
 <name>R1</name>
  <value type = "ltl">
   [] (noChangeAssumptionNextState
        -> O (agreeOnActiveSide \verb+\/+ O (noSideFailed -> agreeOnActiveSide))) .
 </value>
</command>
\end{alltt}
\normalsize

\noindent
New formulas can be defined in the property file
using  the \texttt{definition} tag.
For example, a new constant \texttt{side1Failed}, can be   defined as follows:

\small
\begin{alltt}
<definition>  
 <name>side1Failed</name>
  <value>
    value of side1Failed in component 
      (MAIN -> sideOne -> sideProcess -> sideThread) is true
  </value>  
</definition>
\end{alltt}
\normalsize

%% file: ptolemy.tex
\section{Ptolemy~II Discrete-Event Models}

Ptolemy~II~\cite{ptolemy} 
is a well-established modeling and simulation tool used in industry
that 
provides a powerful yet intuitive graphical modeling
language. 
A Ptolemy~II model consists of a set of interconnected  {\em
  actors} with
\emph{input ports} and \emph{output ports}, where communication channels 
pass \emph{events} from one port to 
another. Such a model 
can be encapsulated as a \emph{composite} actor, 
which may also have input and
output ports.
In Ptolemy~II, real-time systems are modeled as
\emph{discrete-event} (DE) models. Like many  graphical modeling
languages, Ptolemy~II DE models lack formal verification capabilities. 

In each iteration
of a DE model, all components with input execute
\emph{synchronously}. Since connections are  instantaneous and the
components execute in lock-step, we must compute the
\emph{fixed point} of the input for each component in the round before
its execution; this input
comes from the output of another actor's execution in the same
synchronous round.
Formalizing the Ptolemy II DE modeling language is fairly challenging
  as it combines a synchronous fixed-point semantics with time and
  hierarchical models, as well as a powerful expression language.

Real-Time Maude code generation
and verification has been integrated into Ptolemy~II by Kyungmin Bae,
so that a large subset of Ptolemy~II DE models can be verified
from within Ptolemy~II. The paper~\cite{ptolemy-journal} explains the
Real-Time Maude semantics and formal analysis support for Ptolemy~II
DE models in detail.

\begin{figure}[htb]
\begin{center}
\leavevmode
\includegraphics[width=0.85\textwidth]{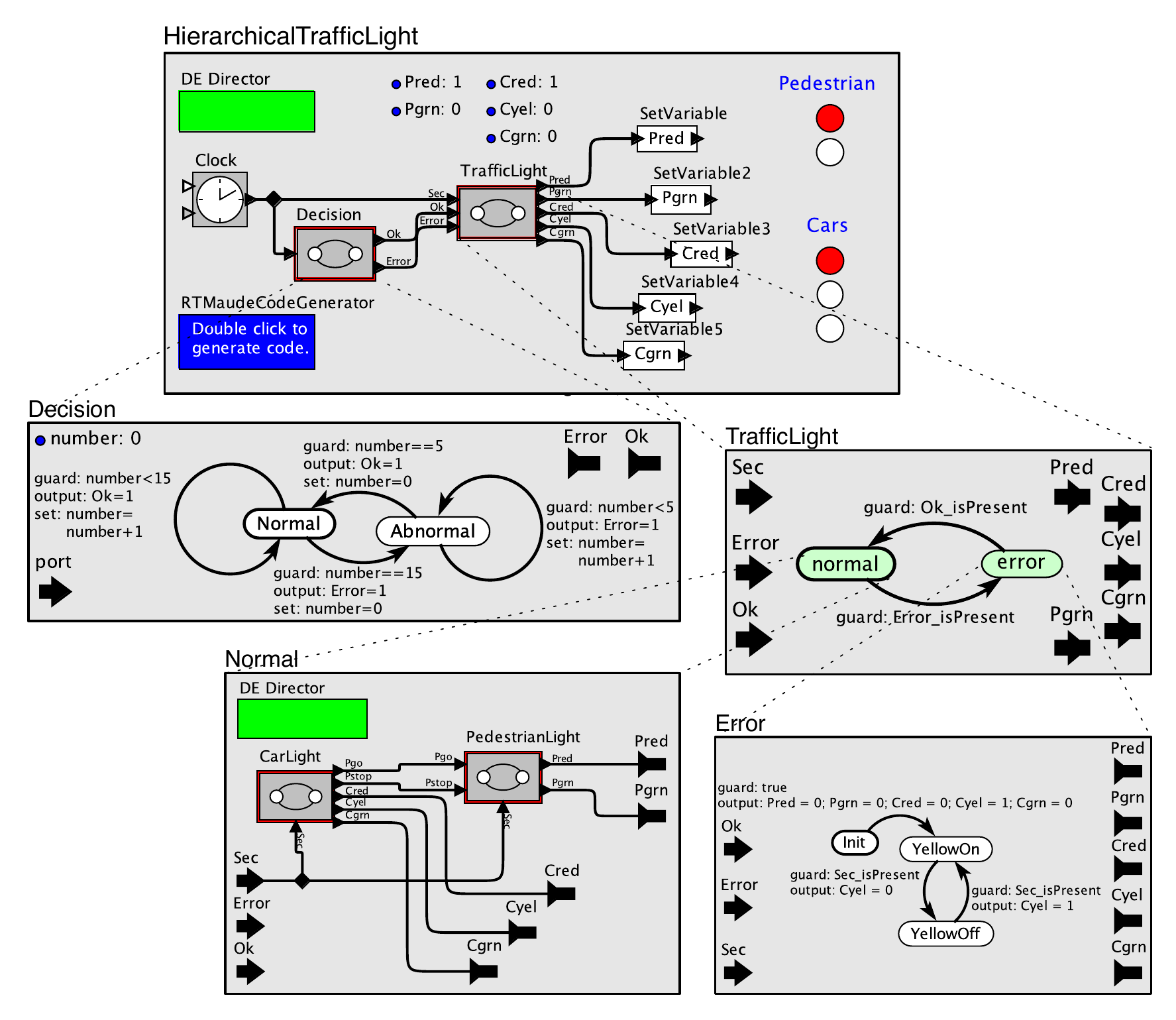}
\end{center}
\caption{A hierarchical fault-tolerant traffic light system in Ptolemy~II.}
\label{fig:ptolemy-htl}
\end{figure}
 
\begin{figure}
  \centering
  \subfloat[\texttt{Pedestrianlight}]{\label{fig:htl-ped}\includegraphics[width=0.3\textwidth]{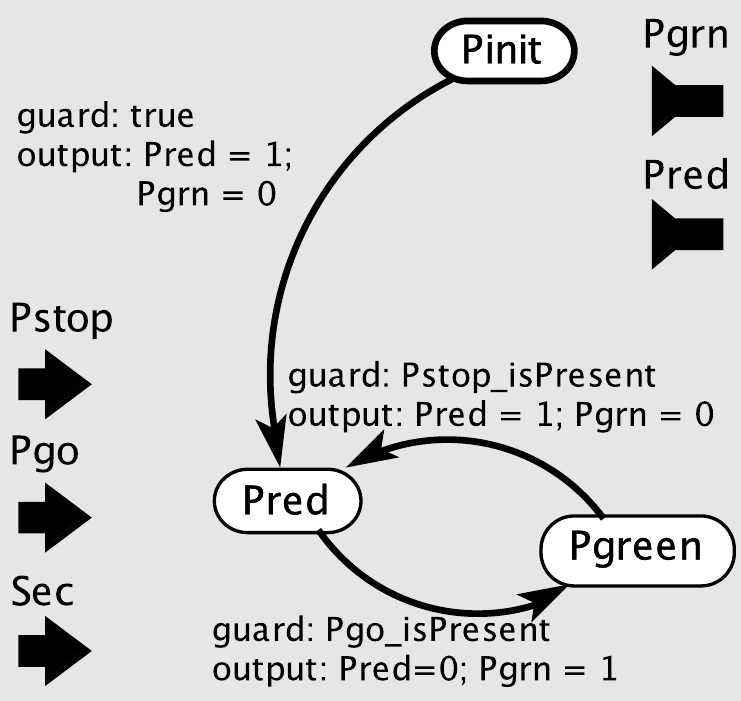}}                
  \hspace{0.02\textwidth}\subfloat[\texttt{Carlight}]{\label{fig:htl-car}\includegraphics[width=0.5\textwidth]{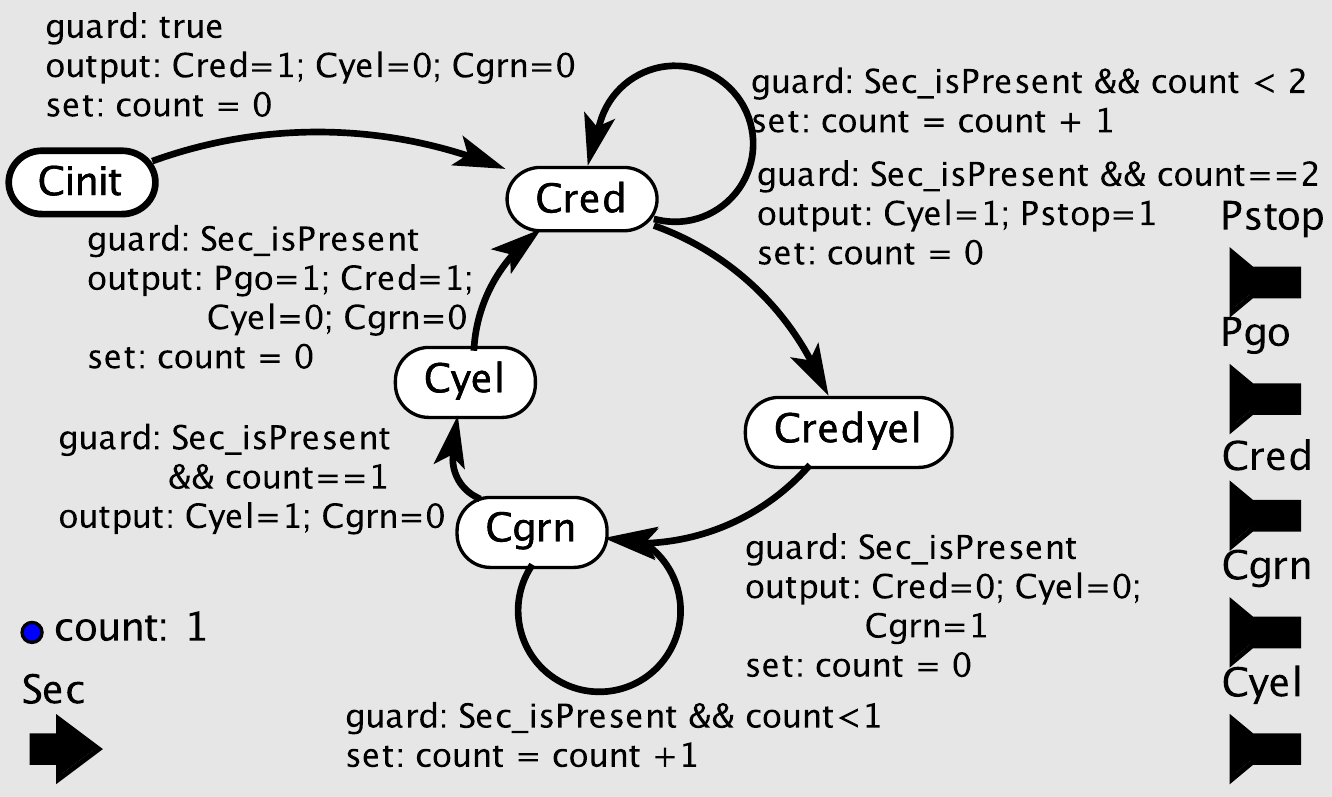}}

  \caption{The state machine  actors for pedestrian lights and car lights}
  \label{fig:htl}
\end{figure}

Figure~\ref{fig:ptolemy-htl} shows a hierarchical Ptolemy~II model of
a fault-tolerant traffic light system at a pedestrian crossing,
consisting of one car light and one pedestrian light. Each light is
represented by a set of \emph{set variable} actors (@Pred@ and @Pgrn@
represent the pedestrian light, and @Cred@, @Cyel@ and @Cgrn@
represent the car light). A light is \emph{on} iff the corresponding
variable has the value 1. The state machine  actor @Decision@
``generates'' failures and repairs by alternating between staying in
location @Normal@ for 15 time units and staying in location @Abnormal@
for 5 time units. 
Whenever the model operates in @error@ mode, all lights are turned
off, except for the yellow light of the car light, which is
blinking. We refer 
to~\cite{ptolemy-journal} for a thorough explanation of the model.

\begin{figure}[htb]
\begin{center}
\leavevmode
\includegraphics[width=0.77\textwidth]{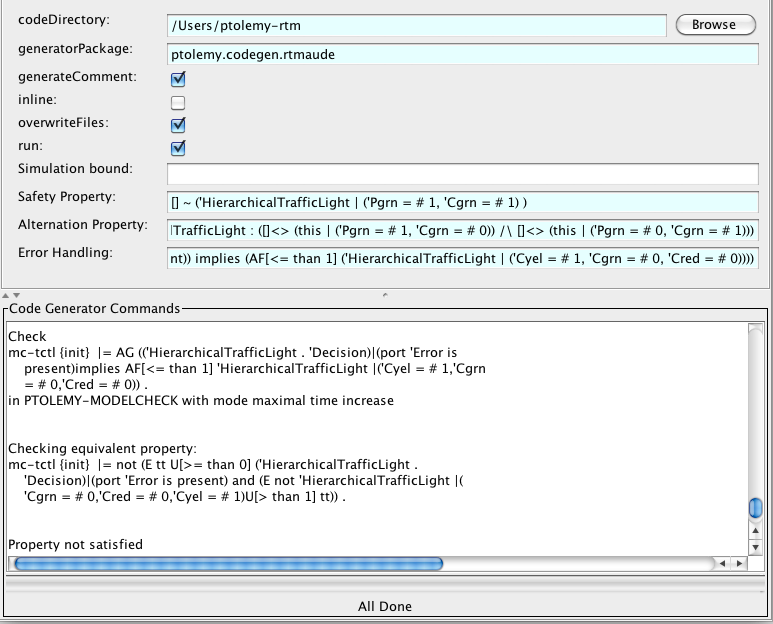}
\end{center}
\caption{Dialog window for the hierarchical traffic light code generation.}
\label{fig:dialog}
\end{figure}

We have used Ptolemy~II's   code generation
infrastructure to integrate both the synthesis of a Real-Time Maude
 model  from a Ptolemy~II design model as well as
Real-Time Maude model checking of the synthesized model into
Ptolemy~II itself.  
When  the blue 
\texttt{RTMaudeCodeGenerator} button in a Ptolemy~II DE model 
is double-clicked, Ptolemy~II
opens a dialog window  (shown in Fig.~\ref{fig:dialog})  which allows the user to start
code generation and to give  model checking 
commands to formally analyze the generated model.
%

We have also predefined in our model checker useful atomic
propositions similar to those in our AADL model checker. 
For example, the proposition
\[ \mathit{actorId}  @ | @ \mathit{var_1}\: @=@ \:
\mathit{value_1}@,@ \ldots @,@\: \mathit{var_n}\: @=@\:
\mathit{value_n}\]   
holds in a state if the value of the  parameter $var_i$ of the  actor
$\mathit{actorId}$  equals $value_i$
for each $1 \leq i \leq n$, where
$\mathit{actorId}$ is the \emph{global actor identifier} of a given actor.
Similarly, 
$\;\;\; \mathit{actorId}\;  \verb+|+\; @port@\; p\; @is@\; \mathit{value}\;\;\;$ and  
 $\;\;\;\mathit{actorId}\;  \verb+|+\; @port@\; p\; @is@\; \mathit{status}\quad$ 
hold if, respectively,  the port $p$ of actor $\mathit{actorId}$ has
the value $\mathit{value}$  and  status $\mathit{status}$.
For state machine actors, the proposition
$\quad\mathit{actorId}  \verb+ @ + \mathit{location}\quad$
is satisfied if and only if the actor $\mathit{actorId}$ is 
in ``local state'' $\mathit{location}$. 

In  the  traffic light system, 
the following \emph{timed} CTL property 
states that the car light will turn yellow, \emph{and only yellow}, 
 within 1 time unit of a failure:

\small
\begin{alltt}
AG (('HierarchicalTrafficLight\!\! .\!\! 'Decision\! |\! port 'Error is present)
   => AF[<= 1] ('HierarchicalTrafficLight\! |\! 'Cyel\! =\! 1, 'Cgrn\! =\! 0, 'Cred\! =\! 0))
\end{alltt}
\normalsize

\noindent
Model checking this property revealed a  previously unknown error:
after a failure, the car light may show red or green in 
addition to blinking yellow.

%% file: moment2.tex
\section{Timed Model Transformations in MOMENT2}

The MOMENT2~\cite{boronat-meseguer-fase08,boronat-meseguer-tools09}
formal model transformation framework is based on a formalization of
MOF  meta-models in rewriting logic. The static semantics of a
system is given as a class diagram (or meta-model) describing the set of
valid system states (or models) that 
are represented as object diagrams,  and the dynamics of a system
is defined as an \emph{in-place model transformation}. In MOMENT2, 
 a model transformation is defined as a set of production rules. 
 Each such  rule 

\small
\begin{alltt}
rl \(l\) \char123 nac \(\mathit{dl}\) \(\mathit{nacl}\) \char123 \(\mathit{NAC}\) \char125 such that \(\mathit{cond}\) ;...
     \!  lhs \char123 \(\mathit{dl}\) \char123 \(L\) \char125 \char125;   rhs \char123 \(\mathit{dl}\) \char123 \(R\) \char125 \char125;   when \(\mathit{cond}\); ... \char125
\end{alltt}
\normalsize

\noindent has a left-hand side $L$, a right-hand side $R$, a set of (possibly
conditional) negative application conditions $\mathit{NAC}$, and a condition with
the \emph{when} clause. $L$, $R$, and $\mathit{NAC}$ contain model patterns, 
where nodes are
object patterns and unidirectional edges are references between objects.
MOMENT2 provides a range of analysis methods in Maude, including a
checking whether a model conforms to its meta-model as well as model
checking of model transformations.

In joint work with Artur Boronat, I have
extended MOMENT2 to add timed behaviors to model transformations
 by  providing a small and simple but  useful set of built-in types for defining
\emph{clocks}, \emph{timers}, and what we call \emph{timed values},
which are clocks that increase with a given rate~\cite{fase10}. 
As a consequence, a software
engineer can use MDA standards and Eclipse Modeling Framework (EMF)
 technology to formally analyze
RTESs.

This approach, where timed constructs have to be added to the original model 
to express timed behaviors requires changing the structural design of the application. 
However, using MOMENT2's
support for \emph{multi-model} transformations, one can
also define a timed system in a \emph{non-intrusive} way, in which the 
user-defined meta-model
of the system is \emph{not} modified. See~\cite{fase10} for details.

The Real-Time Maude semantics of timed model
transformations is a fairly straight-forward extension of  the rewriting logic semantics of model
transformations given in~\cite{boronat-heckel-meseguer-fase09}.
Although MOMENT2 currently uses Maude as its execution engine,
 we can also easily perform Real-Time Maude inspired
\emph{time-bounded} 
analyses by just adding a single unconnected @Timer@---whose initial
value is the time bound---to the initial state.    When this timer expires,
time will
not advance further in the system,  since no rule resets or turns off the timer.
 This ability to perform time-bounded analysis  makes (time-bounded) LTL model
checking analysis possible for systems with an infinite reachable state
space, that can otherwise not be subjected to LTL model
checking.

%% file: emotions.tex
\section{Domain-Specific Visual Languages in e-Motions}

The e-Motions model transformation framework~\cite{emotions}  for
domain-specific visual languages is also based on EMF, with  Ecore
meta-models  defining the abstract syntax of the language, and where
 the concrete syntax maps elements of the abstract syntax onto
graphical objects. Although specification of system behaviors is based
on in-place model transformations,  the approach to timed
behaviors is   different from the one taken in MOMENT2. 
Instead of adding external ``timed constructs,'' e-Motions 
tries to  make the definition of
  timed model transformations as intuitive and high-level as possible
 by providing different kinds of \emph{timed model
  transformation rules}. 

A model transformation rule is equipped with a time interval that 
defines the range of the possible \emph{durations} 
of the rule, which is applied eagerly. A rule can, however,  
be declared to be \emph{soft}, which means that
it is not triggered eagerly, and/or  may be declared  to be 
\emph{periodic}, in which case it is triggered periodically as long it is enabled.  In addition to atomic
rules, there are also \emph{ongoing rules} that 
model continuous actions. 
These are powerful high-level constructs that typically imply that
many different timed rules are being   applied simultaneously to an 
object.

e-Motions has a Real-Time Maude semantics. e-Motions models
can be simulated visually  in the
e-Motions tool, which is available as an Eclipse plug-in. For
reachability and LTL model checking analyses, however, the tool
generates the corresponding Maude model, and the
analysis has to be done at the Maude level. However, the operator \verb+<<_;_>>+ is useful to define
analysis commands without having to know the Maude representation of an
e-Motions model. The term \verb+<<+ \emph{ocl-expr} \verb+;+
\emph{model} \verb+>>+ evaluates the OCL expression \emph{ocl-expr} in
the model \emph{model}. Therefore, to search for a model reachable
from an initial model \verb+myModel+ that satisfies the OCL expression
\verb+ocl-expr+, we  use the Maude command 

\small
\begin{alltt}
search [1] init(myModel) =>* \verb+{+MODEL:@Model\verb+}+ in time T:Time
  such that << ocl-expr ; MODEL:@Model >> .
\end{alltt}
\normalsize

%% file: musab.tex
\section{A Modeling Language for Handset Software}
\label{sec:docomo}

In~\cite{musab-fase09}, Musab AlTurki and researchers at DOCOMO USA
Labs describe a simple but powerful specification language, called
$\mathcal{L}$, that is claimed to be well suited for describing a
spectrum of behaviors of various software systems. The language
provides flexible SDL-inspired timing constructs that yield a more
expressive language for timed behaviors than Erlang~\cite{erlang},
since some nested timing patterns, which can be expressed in
$\mathcal{L}$, are not expressible in Erlang~\cite{musab-fase09}. 

The language has an expression language, imperative features for
describing sequential computations, and asynchronously communicating
processes that can be dynamically created or destroyed. It is worth
remarking that just  the dynamic process creation places $\mathcal{L}$
outside the class of systems that can be represented as timed
automata; so does its expression language and imperative features
which make $\mathcal{L}$ Turing-complete. 

The semantics of of $\mathcal{L}$  is defined in Real-Time Maude, and Real-Time Maude therefore
also provides a prototype analysis tool for $\mathcal{L}$. However, there does not 
seem to be any support for 
automatically translating $\mathcal{L}$ specifications into Real-Time Maude
models. 

%% file: concl.tex
\section{Concluding Remarks}

This paper has motivated the use of Real-Time Maude as a target formalism and tool
that should be well suited to support the formal model engineering of real-time embedded systems,
and has given a brief overview of the use of Real-Time Maude for such purposes. 

Much work remains as always, including covering other languages and larger subsets of the languages mentioned in the paper. 
Furthermore, although we can predefine useful propositions so that the user
can easily define his/her own queries, and although there is a 
one-to-one correspondance between the informal and the formal model,
results from the formal analyses should still be mapped back into the original formalism. 

\paragraph{Acknowledgments.} I would like to thank the organizers of AMMSE 2011 for inviting
me to give a survey talk about the use of Real-Time Maude to support the formal model engineering of embedded systems.
 Much of the work reported in this paper 
is joint work. I am particularly grateful to Jos\'e Meseguer for initiating, encouraging, and collaborating on 
most of  the research presented in this paper in which I have been involved. I am also grateful to Kyungmin Bae for our joint work
on  Synchronous AADL and Ptolemy~II, and to other collaborators on the  projects summarized in this paper,
including Abdullah Al-Nayeem, Artur Boronat, Edward Lee, Steve Miller, Min Young Nam, Lui Sha, and Stavros Tripakis. 
I also gratefully acknowledge financial support from The Research Council of Norway.